\documentclass{iau}

\usepackage{amsmath}
\usepackage{graphicx}
\usepackage{multirow}

\begin{document}

\lefttitle{(Toward) Discovery of Life Beyond Earth and its Impact}
\righttitle{Proceedings Kavli-IAU Symposium No. 387}

\jnlPage{1}{7}
\jnlDoiYr{2021}
\doival{10.1017/xxxxx}

\aopheadtitle{(Toward) Discovery of Life Beyond Earth and its Impact}
\aopheadtitle{Proceedings Kavli-IAU Symposium No. 387}
\editors{H. Landt, M. Dominik \& C. Oliver, eds.}

\title{Venus Phosphine: Updates and lessons learned}

\author{David L Clements}
\affiliation{Imperial College London}

\begin{abstract}
The discovery of phosphine in Venus's atmosphere provides lessons for the search for life. The detection has survived all challenges and has acquired independent support from archival data from PVP. The presence of phosphine in Venus' oxidising environment is perplexing, and comprehensive studies rule out all known abiotic sources. More data is needed to understand the origin of phosphine, leading to JCMT-Venus, a long term atmospheric monitoring programme. This can find how phosphine varies in relation to other species providing clues to its origin. We present the latest JCMT-Venus results. The discovery and subsequent papers were explicit that they did not constitute evidence for life, only of phosphine. Media and public reaction to the discovery and its implications provide lessons for future life searches, as does the reaction of the scientific community. How this was handled by the team, media, and general public will be reviewed.
\end{abstract}


\maketitle

\section{Introduction}

While the surface of Venus is well known to be thoroughly hostile to the presence of life, with temperatures in excess of 450 C and atmospheric pressure nearly a hundred times that of sea level on Earth, there has long been speculation about the possibility of life in the clouds that make up a `temperate zone' of the atmosphere at an altitude of $\sim$ 55 km. At these heights the atmospheric pressure is around 1 bar, and temperatures are around 60 C \citep{m67}. These ideas persist to the present day, fuelled by a number of well established anomalies in various aspects of Venus' atmosphere that remain unexplained \citep{b21}. These include an unidentified UV absorber, the presence of oxygen at $\sim$10 parts per million (ppm) levels in the clouds, the nature of large, 7$\mu$m sized, particles of unknown composition and possibly aspheric shape, and the vertical abundance variations of SO$_2$ and H$_2$O (see \citep{c22} and references therein). Recent modelling of Venus' history, suggesting that it might \citep{w16} (or might not \citep{t21}) have had a warm wet past until as recently as 700 Myr ago, have added to this speculation, with the possibility that pre-existing surface life might have evolved to survive in the cloud decks as the surface became increasingly hostile.

In 2021, a new anomaly was added to the list of issues with our understanding of the atmosphere of Venus with the discovery of phosphine, PH$_3$, above the cloud decks \citep{g21}. This sparked considerable interest for a number of reasons. Firstly, phosphine, as a reduced chemical, should not have been present at all in the oxidising atmosphere of Venus. This is confirmed by detailed investigation of possible chemical pathways for phosphine's formation and destruction on Venus \citep{b21b, b24}. Secondly, phosphine has been proposed as an exoplanet biomarker gas \citep{s20}, suggesting that its presence on Venus might suggest the presence of life. Needless to say, this discovery led to considerable discussion and further work, some of which is detailed in \cite{c22}.

We here present several updates on our observations of phosphine in the atmosphere of Venus, and especially the results of the JCMT-Venus long term monitoring programme. We also, in the spirit of this meeting preparing us for the discovery of life elsewhere, present a discussion on how the initial phosphine results were received by the scientific community and the general public, and what this might mean for the reception of future discoveries in astrobiology.

\section{Current Status of the Phosphine Detection}

The first announcement of the discovery of phosphine in Venus' atmosphere \citep{g21} was a surprise and, understandably, was challenged on a number of fronts. These include the suggestions that the weak absorption line that indicated the presence of phosphine was a data processing artefact and not in fact real \citep{v21,to21,s20}. These suggestions were subsequently shown to be incorrect \citep{g21b} since the chances of an artefactual line appearing at the exact frequency of a known absorption line in these datasets is $\sim 1\%$, and the chances of a line appearing at the same frequency in {\em both} datasets is far smaller. It was also suggested that the detected line, while real, was not phosphine but was in fact a misidentified SO$_2$ line \citep{l21}. This cannot be the case for two reasons: firstly, the central line frequency of the SO$_2$ is inconsistent with the measured line centre by $\sim 3 \sigma$, and, secondly, because contemporaneous or near-contemporaneous observations of a different SO$_2$ line limit the maximum contamination of phosphine to be $< 10\%$. Meanwhile, reanalysis of the Pioneer Venus Probe mass spectrometer {\em in situ} data from 1978 \citep{m21} suggest the presence of phosphine at part-per-million (ppm) levels within the clouds, while the ALMA and JCMT data indicate 10-20 part-per-billion (ppb) levels above the clouds.

Other observations in search of phosphine have also been conducted, producing a mixed set of upper limits and detections \citep{c22, g23}, but a picture may be emerging with detections of phosphine appearing as the atmosphere as observed moves from night into day, while upper limits or weak detections are seen as it moves from day into night, suggestive of phosphine being destroyed by solar photolysis \citep{b21b} during daylight.

\section{The Next Step: JCMT-Venus}

One thing that is clear from the discussion of the \cite{g21} discovery of phosphine is that more and better data are needed to both confirm the detection and to begin investigations of how phosphine abundances might vary in relation to other properties, such as time of day or abundances of other chemical species such as H$_2$O or SO$_2$. The JCMT-Venus project was thus conceived \citep{c24}. This is a long term JCMT programme to monitor the atmosphere of Venus at mm wavelengths using the new '\={U}'\={u} receiver on the N\-{a}makanui instrument \citep{m20}. The previous receiver used, RxA3, was retired from the telescope in 2020, and the new instrument has a number of significant improvements relevant to this project. These include improved sensitivity and, more importantly, a significantly increased bandwidth, which allows phosphine, HDO, as surrogate for H$_2$O, and SO$_2$ lines at mm wavelengths to all be observed simultaneously. JCMT-Venus was awarded a total of 200 hours of telescope time to monitor Venus during three campaigns in Feb 2022, July 2023 and September 2023. So far about 100 hours of this time has been used.

\begin{figure}
    \centering
    \includegraphics[width=14cm]{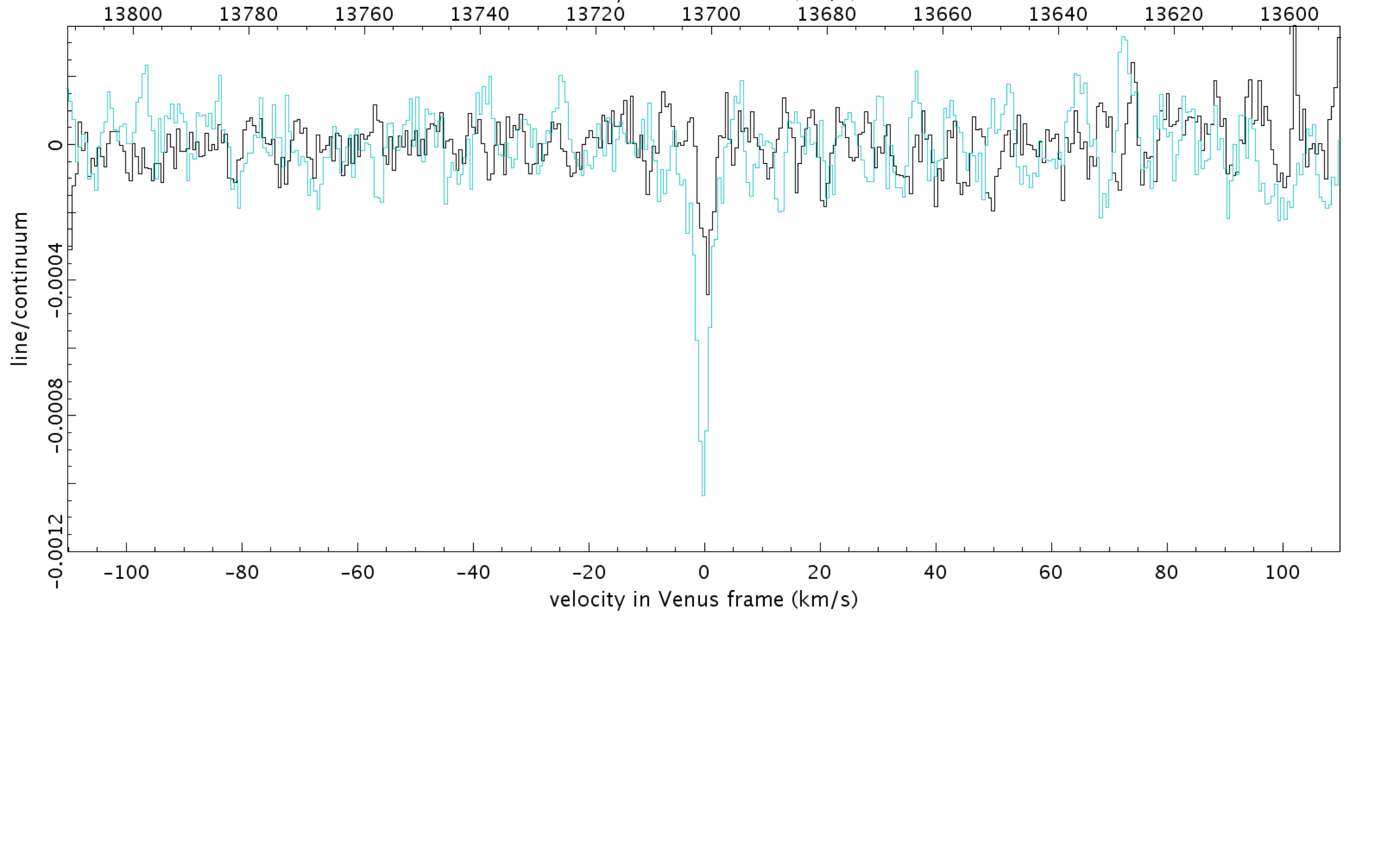}
    \caption{The result of applying the fourier method to remove ripples to the first few days of the first JCMT-Venus campaign. As can be seen both the HDO line (shown in blue) and the phosphine line (shown in black) are recovered.}
    \label{fig1}
\end{figure}

The results of these observations are currently being analysed by the JCMT-Venus team. The data are not simple to deal with since the extreme brightness of Venus as a millimetre source, compared to the usual targets of the JCMT, leads to numerous systematic effects becoming significant. These are principally due to reflections of Venus from the telescope, telescope dome and support structure, and within the receiver cabin, entering the beam of the instrument. The digital autocorrelation analysis used by the instrument then leads these time delayed signals to appear in the resulting spectrum as various sinusoidal waves which contaminate the data in the form of large scale, time-varying baseline drifts and sinusoidal ripples with amplitudes exceeding those of the expected absorption lines. The large scale drifts can be dealt with by the subtraction of a running baseline average, though this comes at the cost of suppressing the broad wings of any absorption lines, which would originate from the clouds decks themselves thanks to pressure broadening. The remaining baseline ripples would be dealt with in narrow band instruments by fitting and removing a polynomial baseline from the data, with the positions of any expected absorption lines masked and interpolated over (see eg. \cite{g21}). For the broader spectral range in this data, which include multiple peaks and troughs from these sinusoidal ripples, such an approach would require a very high order polynomial, and so a different, Fourier based, approach may be needed \citep{g24}, whereby the data is Fourier transformed, all but the most significant frequency components, representing the baseline ripples, are removed. This results in a `ripple spectrum' which is transformed back and removed from the original data, suppressing the reflection-induced systematics. The result of applying this technique to the first few days of JCMT-Venus data for both the HDO and phosphine lines are shown in Figure \ref{fig1}. As can be seen both lines are recovered with good signal to noise.

While this approach allows us to retrieve the narrow components at the line cores, as can be seen in Figure \ref{fig1}, the broad line wings, coming from deeper in the atmosphere, remain suppressed. These broad wings are important as they allow us to assess the abundances of species deeper into the atmosphere. To investigate these broad line wings a new approach has been devised which relies on the repeatability of the large baseline ripples across our spectra to perform what is effectively {\em post-hoc} frequency switching. This removes the repeatable element of the large scale baseline ripples while retaining the broad wings of any absorption lines that might be present. While this method remains experimental, and might not be applicable to all of our data, the initial results are very encouraging, revealing broad wings in the phosphine line while at the same time recovering broad line wings in HDO which are consistent with the line shape and depth of HDO seen in near-simultaneous observations from the IRTF \citep{e23}. The phosphine abundance at at altitude of $\sim$ 55 km derived from this preliminary analysis is about 0.3 ppm, which is broadly consistent with the levels seen at similar altitudes by the PVP mass spectrometer \citep{m21}.

Much more work remains to be done to fully process and understand the JCMT-Venus data - the first campaign of which on its own produced 140 times as much information on Venus' atmosphere than the data used for the initial phosphine discovery. However, it is already clear that the data has already confirmed the presence of phosphine and will provide many new insights once the instrumental systematics are fully understood and removed.

\section{Lessons Learned}

While the origin of phosphine on Venus remains unclear, a biological origin cannot be discounted since no satisfactory standard chemical origin has yet been found \citep{b21}. Our original discovery of phosphine in 2020 thus had a significant impact as a potential sign that life might exist elsewhere. As such, its reception, both scientifically and more broadly, can serve as a case study for what might happen when there is a clear detection of life elsewhere, potentially through the examination of the chemical composition of an exoplanet's atmosphere. We here discuss the scientific and public reaction to this result, and the lessons that might be learned.

\subsection{Scientific Reception}

As is appropriate for any potentially significant result, the scientific community rigorously tested our work (see discussion in Section 2). This was facilitated by making our data publicly available to any who wished to examine it. This is the first lesson that should apply to future astrobiology results: {\bf the data used to derive any significant results should be made available to the astronomical community so that the results can be independently verified.}

Acceptance of our result is still far from universal, and this is likely to, at least initially, be the case for any detection of signs of life elsewhere. Additional observations, in our case from JCMT-Venus, will be critical to improve the robustness and understanding of any claimed life detections. There is a risk that the reluctance of the community to accept a radically new result may be an impediment to getting the additional telescope time to confirm and better understand such a result. We have certainly had this difficulty at some telescopes (though clearly not the JCMT). The second lesson learned is thus: {\bf there need to be clear routes to obtaining confirmatory observations for claimed astrobiological detections even in the presence of skepticism by the community}. Given the highly oversubscribed nature of world ranking telescopes, and the resulting extreme conservatism of time assignment committees, this may mean that directors discretionary time is needed. Good communication with relevant telescope directors in advance of any detection will help this process.

In the case of Venusian phosphine we have the added advantage that ground-truth observations from flyby, orbiting, or descent probe missions that can provide fundamentally better and less ambiguous data than is possible for observations from Earth. In the long term, sample return missions will allow detailed biochemical analysis of any life found in the Solar System, answering fundamental questions about the origin and evolution of life. Such detailed biochemical analysis will never be possible for observations of exoplanets. This leads to the possibility that alternative, non-biological explanations for any claimed detection can never be fully excluded, and our third lesson learned: that {\bf some people will never be convinced}.

\subsection{Public Reception}

We made detailed plans for public outreach in advance of the publication of the \cite{g21} paper, which was accompanied by a press conference and various media releases, some of which were recorded months before. We also, with the help of the EAO, arranged for the media impact to be monitored so that we could quantitatively assess the reception. The whole team was briefed that at all times we should make it clear that biological activity was only one possible explanation for the presence of phosphine in the atmosphere of Venus. This was, in contrast to the ALH84001 meteorite result, to avoid over-hyping the result and hopefully prevent any extreme reactions to the result. This was largely successful, though at least one journalist thought we were suggesting that penguins (whose guano is a known source of phosphine on Earth) were aliens originating from Venus (they were rapidly disabused of this notion when contacting team members).

There was considerable media interest, with a total aggregate readership of over 15 billion, making it one of the most impactful scientific announcements that year. The clear lesson to learn from this is to: {\bf prepare for substantial media interest, but to emphasise that uncertainties in the conclusions remain, and to not engage with those who have extreme or unreasonable reactions.} It might also be useful to reflect on the political involvement with the ALH84001 result and how that led to the over-hyping of a result that had not yet been fully refereed in the standard publication process. We thus also conclude that: {\bf publication by press release is a bad idea.}

The overall impact of the detection of phosphine on Venus is analysed from a rather different perspective, including social and psychological as well as scientific factors, elsewhere in this volume \citep{v24}.

\section{Conclusions}

Our discovery of phosphine in the atmosphere of Venus has so far survived detailed scientific scrutiny, and new observations from the JCMT-Venus project have confirmed its presence and have the potential to help us understand its origin and significance. New techniques are necessary to deal with systematic problems with the new data, and application of these leads to not only confirmation of phosphine, but detection of its pressure broadened line wings indicating amounts of phosphine in the clouds consistent with the ppm levels seen in PVP {\em in situ} mass spectrometer data. The reception of our original phosphine results provides lessons for the release of future exobiological results, including the need to make the data accessible, the need to be able to acquire followup observations, and the need to prepare an appropriately calm and consistent response to press and public interest.

\acknowledgements Many thanks to Jane Greaves and Wei Tang for numerous contributions and discussions, and to the whole JCMT-Venus team without whom this work would not have been possible. The JCMT is operated by the East Asian Observatory on behalf of The National Astronomical Observatory of Japan; Academia Sinica Institute of Astronomy and Astrophysics; the Korea Astronomy and Space Science Institute; the National Astronomical Research Institute of Thailand; Center for Astronomical Mega-Science (as well as the National Key R\&D Program of China with No. 2017YFA0402700). Additional funding support is provided by the Science and Technology Facilities Council of the United Kingdom and participating universities and organizations in the United Kingdom and Canada.

\end{document}